\documentclass[twocolumn]{revtex4}
\usepackage{graphicx}
\usepackage{amsmath,amssymb}
\usepackage{epstopdf}
\usepackage{bm}
\newcommand{\bra}[1]{\langle #1 |}
\newcommand{\ket}[1]{| #1 \rangle}

\renewcommand{\vec}[1]{{\bm #1}}
\newcommand{\br}{\mathbf r}
\newcommand{\bR}{\mathbf R}

\begin{document}

\title{Ab-initio path integral techniques for molecules}
\author{Daejin Shin}
\author{Ming-Chak Ho}
\altaffiliation[Present address: ]{Deptartment of Physics and Astronomy, University of Southern California, Los Angeles, California 90089-0484.}
\author{J. Shumway}
\email{shumway@mailaps.org}
\affiliation{Department of Physics, Arizona State University, Tempe, AZ 
85287-1504}\date{\today}       % Activate to display a given date or no date
\begin{abstract}
Path integral Monte Carlo with Green's function analysis allows the sampling of 
quantum mechanical properties of molecules at finite temperature.
While a high-precision computation of the energy 
of the Born-Oppenheimer surface from path integral Monte Carlo is quite costly,
we can extract many properties without explicitly calculating the electronic 
energies. We demonstrate how physically relevant quantities, such as 
bond-length, vibrational spectra, and polarizabilities of molecules may be 
sampled directly from the path integral simulation using Matsubura (temperature) 
Green's functions (imaginary-time correlation functions).
These calculations on the hydrogen molecule (H$_2$) are a proof-of-concept, 
designed to motivate new work on fixed-node path-integral calculations for molecules.
\end{abstract}
\maketitle

\section{Introduction}
The quantum Monte Carlo (QMC) method has been extensively used as a powerful 
technique for calculating some physical properties of many-body systems \cite{Guardiola:1998},
such as molecules \cite{Reynolds:1982, Hammond:1994}, solids \cite{Foulkes:2001},
 and nanostructures \cite{Shumway:2006b}.
For molecules, it has been reported that the diffusion QMC method can achieve
near chemical accuracy  within the fixed-node approximation \cite{Grossman:2002,Benedek:2006}.

While variational and diffusion QMC algorithms usually sample only electrons at zero temperature, 
path integral Monte Carlo (PIMC) can sample both the ions and electrons
 simultaneously at finite temperature.
This is a promising approach to calculate thermal effects in the molecular systems. 
Additionally, PIMC includes the zero-point motion of the ions, 
yeilding the correct quantum statistical treatment of the motion of light ions at all temperatures.

We note that another approach, coupled electron-ionic Monte Carlo 
(CEIMC) \cite{Ceperley:2002,Pierleoni:2006}, has been demonstrated in which 
electrons and ions are sampled in two independent but coupled 
Monte Carlo algorithms. For the molecule we study in this paper, a comparable CEIMC simulation
would have a ground-state variational or diffusion QMC simulation of the electrons that determines 
the Born-Oppenheimer energy surface. These QMC electron energies would then be used in a path 
integral simulation for the protons. Because the QMC estimates of ground-state electronic energies 
have statistical error bars, a penalty-method \cite{Ceperley:2002} would be used to partially compensate
for non-linear biases when these energies are fed into the ionic path integral Monte Carlo.
Ultimately, the CEIMC method requires a fairly accurate determination of total energies at many points
on the Born-Oppenheimer surface for many different configurations of the ion positions.

In this paper we address the following issue: by treating electrons and ions in the same path
integral simulation (avoiding the explicit calculation of the Born-Oppenheimer energy surface), can
we extract properties of a molecule, such as bond-lengths, vibrational frequencies, and
polarizability? These properties are often calculated directly from the Born-Openheimer
surface, i.e. bond-lengths and vibrational frequencies come from local minima and curvature
of the energy surface. Here we treat these properties as expectation values of the electron-ion thermal
density matrix. That is, we do not make the Born-Oppenheimer approximation, instead, we
sample the electron and ion coordinates in a single path integral. The calculations 
presented here are a proof-of-concept, demonstrating how to extract physical properties from
the PIMC simulations. The question of the relative efficiencies of this approach and CEIMC
will have to be left to future calculations on larger systems, most likely requiring some kind of a path-integral fixed-node approximation \cite{Shumway:2005d}.

In PIMC simulations
we can collect various correlation function of the system in imaginary time.  
These functions, known as finite temperature Matsubura Green's functions, 
can be transformed from the imaginary-time domain into the imaginary-frequencies 
through the fast Fourier transform (FFT).
Since we want to calculate the physical properties of the systems,  
the correlation function at real frequency need to be obtained from the imaginary-frequency function.
It is possible through the analytic continuation techniques to obtain real-time correlation function from imaginary-time correlation function,
since a time-ordered correlation function is an analytic function of the time variable 
in the complex plane \cite{Baym:1961}. In other words, the time-correlation function calculated 
along the imaginary axis can be uniquely analytically continued to the real-time axis.
 
For the case of the simple and closed correlation function,  
the analytic continuation of imaginary-frequency correlation function 
is simply the replacement of $i\omega_n$ with $\omega+i\eta$.
But for realistic correlation functions in molecules it is hard to find the closed form of  
imaginary-frequency correlation function.
Therefore analytic continuations need to be performed with numerical methods. 

However, the analytic continuation of numerically evaluated correlation functions
often leads to grossly amplified statistical errors in the real time correlation function.
Some numerical methods have been proposed for reducing the statistical errors,
such as using Pad\'{e} approximations \cite{Thirumalai:1983} or least-squares 
fitting \cite{Schuttler:1985}, minimizing a suitably defined potential \cite{Jarrell:1989}, 
maximum-entropy \cite{Gallicchio:1996, Krilov:2001}, and 
singular value decomposition (SVD) methods \cite{Gallicchio:1998}.
It has been reported that 
the rotational constant of a molecule can be obtained by fitting imaginary-time 
correlation function from PIMC data to 
analytic models of correlation function \cite{Blinov:2004}.

In this paper we show that some properties of molecules can be sampled 
directly from path integral simulations using Matsubura Green's functions. 
For an accurate treatment of the Coulomb interactions between particles
we use a new method to tabulate of the Coulomb density matrix that we have developed.
Using correlation functions sampled with the accurate Coulomb propagator,
we demonstrate PIMC calculations of the bond length, vibrational frequency, and polarizability of a hydrogen molecule. This calculations are important benchmarks for future calculations
on more complicated systems. In particular, we have chosen a simple system without
a fermion sign-problem so that we can separate the analysis of correlation functions
from questions about the fixed-node approximation.

\section{Monte Carlo Sampling of Path Integrals}
\subsection{Monte Carlo Sampling}
In thermal equilibrium, the average value of any physical quantity 
can be calculated from the density matrix.
In the coordinate representation, the thermal density matrix of $N$ particles is defined by 
\begin{equation}
\rho(\bR,\bR';\beta)=\frac{1}{Z}\bra \bR e^{-\beta H}  \ket {\bR'} ,
\end{equation}
where $ \beta =  1/k_{B}T$ is the inverse temperature, 
$\bR =( \mathbf{r}_1 ,\cdots, \mathbf{r}_N )$ are the particle coordinates, 
$\mathbf{r}_i $ is the position of the $i$-th particle,
and $Z= \int\rho(\bR,\bR;\beta)d\bR$ is the partition function.

The average value of any physical quantity $\mathcal O$ can be written 
\begin{equation}
\langle \mathcal{O} \rangle=\int d\bR d\bR'\rho(\bR,\bR';\beta)
\langle \bR|\mathcal{O}| \bR'\rangle  .
\end{equation}
In the path-integral formula, in $N$ slices, we expand the density 
matrix \cite{Feynman:1972},
\begin{equation}\label{eq:primative}
 \begin{split}
 &\rho(\bR_0,\bR_N;\beta)  = \int d\bR_{1}d\bR_{2} \cdots d\bR_{N-1}
    \Bigg( \frac{2\pi\hbar^2\Delta \tau}{m}  \Bigg)^{-3/2} \\
 & \times \exp \bigg[ -\sum_{n=1}^{N} \bigg( 
  \frac{(\bR_{n-1}-\bR_n)^2}{2\hbar^2\Delta \tau/m}   +\Delta\tau V(\bR_n)  \bigg) \bigg] ,
 \end{split} 
\end{equation} 
where  $m$ is the mass of the particle and the time step is $\Delta \tau=\beta/N$. 
In our molecule simulations $\Delta \tau=0.01 \text{ Ha}^{-1}$.
We estimate this high dimensional integral with Metropolis Monte Carlo algorithm,  
with an efficient multilevel sampling method. 
For a detailed discussion of PIMC methods, see reference \cite{Ceperley:1995}.

\subsection{Accurate Coulomb Action}
The primitive approximation to the path integral, Eq.~(\ref{eq:primative}), is
not correct for attractive Coulomb interactions. Instead, we use an improved
quantum action $U(\bR,\bR')$ in place of the bare Coulomb potential $V(\bR)$.
The use of improved actions is discussed in reference \cite{Ceperley:1995}.

In order to more accurately describe the Coulomb interactions 
we have developed a new technique for tabulation of imaginary-time 
Coulomb Green's function \cite{Shumway:2006}, from which we extract the
Coulomb action.
The Green's function, $G(\br,\br')$ for two particles in $N$-dimensional space
satisfies the equation,
\begin{equation}
\left[-\frac{1}{2\mu}\nabla^2+\frac{z}{r} \right]G^{(N)}(\br,\br';\tau) = 
-\frac{d}{d\tau}G^{(N)}(\br,\br';\tau).
\end{equation}
The initial condition is $G^{(N)}(\br,\br';0)=\delta^N(\br-\br')$ and
 $\br$ and $\br'$ refer to the relative separation of two particles with
reduced mass $\mu$, and $z$ is the product of the charges.
There are several ways to numerically evaluate Green's function, 
see reference \cite{Shumway:2006}.
But each methods has its particular limitation 
and numerical errors are often more than one percent. 

Since the Coulomb potential, $V(\vec r) \varpropto \frac{1}{r}$, has 
the symmetry properties, without a partial wave expansion we are able to 
simple, fast, and highly accurate numerical evaluation of $G(\br,\br';\tau)$ 
using one-dimensional FFT's.
The technique relies on Hostler's recursion relation \cite{Hostler:1963, Hostler:1970},
which relates the $N$-dimensional Green's function to the 1-dimensional
function. 
The use of an accurate Coulomb action is especially important in chemistry.
An analogy can be drawn to the use of accurate integration schemes in molecular dynamics.
In the current work, we find that quantities, such as the molecular binding energy and electrical polarizability are especially sensitive to the quality of the tabulated Coulomb action.

\section{Estimators}
\subsection{Energy and the Virial Estimator}
There are several estimators to obtain the total thermal energy of a system.
The most widely used energy estimator is the thermodynamic energy estimator, 
which is obtained by differentiating the partition 
function with respect to $\beta$,
\begin{equation}
E_T=-\frac{1}{Z} \frac{dZ}{d\beta}= \left\langle \frac{dU_i}{d\tau} \right\rangle,
\end{equation}
where $Z$ is the partition function and $U_i$ is the action.
In other words, the thermal energy is an average value of the 
imaginary-time derivative of the action.

For the calculation of molecules we used the virial energy estimator.
The virial estimator of the energy is 
\begin{equation}
E_V = \left\langle \frac{dU_i}{d\tau}-\frac{1}{2}F_i \Delta_i \right\rangle,
\end{equation}
where $\Delta_i$ is the deviation of the $i$-th particle from its average position,
and $F_i$ is a generalization of the classical force,
\begin{equation}
F_i=-\frac{1}{\tau}\nabla_i \Big( U_{i+1}+U_i \Big). 
\end{equation}
For more detailed derivation, see references \cite{Ceperley:1995} and \cite{Herman:1982}.

\subsection{Static polarization}
The general description of the static polarization of molecules is 
$\mathbf{P}=\alpha_{\mu\nu} \mathbf{E} $,
where $\mathbf{E}$ is the applied electric field, 
$\alpha_{\mu\nu} $ is the static polarizability tensor of the molecule, and
$\mu$ and $\nu$ denote vector directions.
For diatomic molecules, symmetry reduces the polarizability tensor to reduced into two components,
the parallel ($\alpha_\parallel$) and perpendicular ($\alpha_\perp$) polarizability.

Using the polarization estimator 
\begin{equation}
\label{eq:statpol}
 \alpha_{\mu \nu}= \frac{1}{E_\nu}\sum_i \left\langle e X_{\mu}(i) \right\rangle ,
\end{equation}
we calculate the static polarizability of the molecules given the electric field.
Here $X_{\mu}(i)$ is the position of the $i$-th particle in the direction of $\mu$,
and $E_\nu$ is the magnitude of the the electric field in the $\nu$ direction.

\subsection{The Matsubura Green's function method}
The Matsubura Green's function method is a very useful technique for
calculating the physical properties of 
many-body systems at finite temperature \cite{Mahan:2000}. 
The general definition of temperature Green's function is
\begin{equation}
G(\tau,\tau') \equiv - \langle T_\tau \mathbf A(\tau) \mathbf B(\tau') \rangle _\beta ,
\end{equation}
where $T_\tau$ is the time-ordering operator,  $\tau$ represent the imaginary time,
and $\mathbf A$ and $\mathbf B$ are operators.
The bracket means the thermal average value at the inverse temperature $\beta=1/k_BT$.
If the Hamiltonian of the system is independent of time,
Green's function becomes only a function of the time difference $\tau-\tau'$,
so it also can be written
$G(\tau)= - \langle T_\tau \mathbf A(\tau) \mathbf B(0) \rangle _\beta$.

A Fourier transform gives the frequency dependence of Green's function 
for bosonic quantities, such as molecular vibrations and polarizabilites,
\begin{equation}
\label{eq:fft}
 G(i\omega_n)=\frac{1}{\beta} \int _0^\beta d\tau e^{i\omega_n \tau} G(\tau),
\end{equation}
where $\omega_n=2n\pi/\beta$ and $n=0, \pm1, \pm2, \cdots$.
In the linear response theory,
the linear response of the system to a small perturbation can be written in terms of the Green's function.
  
As mentioned in the introduction section, through analytical continuation, 
Matsubura Green's function of complex frequencies, $i\omega_n$, 
can be analytically continued to  real, time-ordered correlation 
functions, which determine physical properties of the dynamic system. 
However, in numerical analytic continuation 
even small noise on imaginary-time correlation can make big statistical errors in
real time dynamics of the system \cite{Gallicchio:1998}. 
Instead of that, we analyze directly the imaginary-time correlation function
calculated by path integral Monte Carlo method in order to obtain some properties of molecules.
To calculate bond-length and vibrational frequency,
we use displacement-displacement imaginary-time correlation functions and their Fourier transforms.
For polarizability of molecules we use polarization-polarization correlation function.

\subsubsection{Bond-length and vibrational frequency}
\label{sec:vibration}
For calculating of the bond-length and vibrational frequencies of 
molecules we use the displacement-displacement correlation function.
The displacement correlation function of a harmonic oscillator, 
with $V(x)=\frac{1}{2}m\omega_{\mathtt{SHO}} x^2$, 
in the imaginary-time domain at the inverse temperature $\beta$ can be written
\begin{equation} \label{eq:phononGF}
G(\tau)= - \langle T_\tau x(\tau)x(0) \rangle _\beta , 
\end{equation}
where $x(\tau)$ refers to the displacement of the oscillator from equilibrium.
For $\tau>0$, this correlation function is \cite{Doniach:1974},
\begin{equation}
\label{eq:exactGF}
G(\tau)= -\frac{1}{2m\omega_{\mathtt{SHO}}}\Big[(N+1)e^{-\omega_{\mathtt{SHO}}\tau}
+Ne^{\omega_{\mathtt{SHO}}\tau}\Big],  
\end{equation}
 where $ N= 1/(e^{\omega_{\mathtt{SHO}}\beta}-1)$ is the distribution function 
at the temperature, $k_BT=1/\beta$ and we set $ \hbar=1$.

For the hydrogen molecule, the displacement $x(\tau)$ represents the deviation from
the equilibrium separation distance between two protons. 
The mass, $m$ is the reduced mass of two protons.
At low temperature, we assume that the molecule is in harmonic motion,
so we compare our results with the simple harmonic oscillator.

We are able to extract the bond length from imaginary-time displacement-displacement
correlation function, Eq.~(\ref{eq:phononGF}). 
The function $G(\tau)= - \langle x(\tau)x(0) \rangle$ can measure the fluctuation in the quantity $x$.
At $\tau=0$, the initial value of $G(0)$ is the equilibrium average $-\langle x^2 \rangle $.
As time goes on, the correlation function shows the time decay which measure 
how $x(\tau)$ and $x(0)$ are correlated each other. 
In the long time limit, the correlation function becomes totally uncorrelated 
so that the correlation function can be written as,
\begin{equation}
G(\tau \rightarrow \infty)=-\langle x(\tau) \rangle \langle x(0) \rangle
\end{equation}
At $\tau=\beta/2$, $G(\beta/2) $ is approximately $\langle D \rangle ^2$, 
where $D$ is the bond length of the molecule at low temperature.
Since the bond length can be directly computed as a time-independent average, 
this limit is an important sanity check on our imaginary-time data.

To obtain the vibrational frequency of molecules 
we use imaginary-frequency correlation function.
Using Eq.~(\ref{eq:fft}), we can obtain the correlation function of the simple harmonic oscillator
 in the imaginary-frequency space, 
\begin{equation}
   \begin{aligned} \label{eq:blomega}
G(i\omega_n) =&\frac{1}{\beta}\bigg[ \frac{1/m}{(i\omega_n)^2-\omega_{\mathtt{SHO}}^2} \bigg ] \\
             =&\frac{1}{\beta}\bigg[ \frac{-1/m}{(\omega_n)^2+\omega_{\mathtt{SHO}}^2} \bigg ]  
   \end{aligned}
\end{equation}	  
Once we have the correlation function,
we can calculate the vibrational frequency $\omega_{{\mathtt{SHO}}}$ at $i\omega_n = 0$,
\begin{equation}
 \omega_{{\mathtt{SHO}}}^2 =-\frac{1/(\beta m)}{G(i\omega_n=0)}
\end{equation}
 at each temperatures $\beta$.

Another way to find the frequency is to directly use the correlation function.
Since the correlation function $G(i\omega_n)$ forms a Lorentzian in the imaginary 
axis of complex Green's function, the full width at half maximum (FWHM) of the correlation 
function is equal to the frequency of the harmonic oscillator.
Therefore we can compare to these frequencies using two methods.

\subsubsection{Polarization}
We use the polarization-polarization correlation function to calculate the 
polarizability of molecules. 
The imaginary-time correlation function can be written
\begin{equation}
\alpha_{\mu \nu}(\tau)= - \langle T_\tau P_\mu(\tau)P_\nu(0) \rangle , 
\end{equation}
where $ P_\mu(\tau) = ex_\mu(\tau)$.
Once we have obtained the imaginary-time correlation function,
by Fourier transform Eq.~(\ref{eq:fft}) we can calculated the dynamic polarizability. 

Similar to the bond-length calculation, first we calculate the polarizability for a
simple harmonic oscillator.
(Note that this oscillator is a model for our polarization analysis and is not related to molecular vibrations.) 
From linear responses theory the static susceptibility $\alpha$, 
due to the small electric field perturbation 
$ -exE $, can be written with the first order perturbation theory,
\begin{equation}
\label{eq:SHOpolfq}
\alpha = \frac{2e^2\langle0|x^2|0\rangle}{\omega_{\mathtt{SHO}}}
=\frac{e^2}{m\omega_{\mathtt{SHO}}^2}
\end{equation}
It is clear that the polarizability depends on the confinement potential 
of the harmonic oscillator, $\omega_{\mathtt{SHO}}$.

We can also obtain the polarizability from the imaginary-frequency 
polarization-polarization correlation function.
The correlation function for the simple harmonic oscillator is 
\begin{equation}
\label{eq:SHOpol}
\begin{split}
\alpha(i\omega_n) &= \int _0^\beta d\tau e^{i\omega_n \tau}
  \langle T_\tau P_\mu(\tau)P_\nu(0) \rangle  \\
	&= \frac{e^2/m}{(i\omega_n)^2-\omega_{\mathtt{SHO}}^2} .
\end{split}
\end{equation}

Similar to bond length estimator, we assume that the molecules are in the 
harmonic motion. Once we obtain $\alpha(i\omega_n)$, the polarizability 
$\alpha(0)$ is the value at $i\omega_n =0$. Note that for molecules the 
polarizations $P$ are to be added for all particles in the molecule.
 
\section{Results}
In this section we report the results of our tests on an H$_2$ molecule.
Simulations were performed in serial and in parallel jobs with up to
eight processors using our open-source {\tt pi} code available online at 
{\text http://phy.asu.edu/shumway/codes/pi.html}.
We use a time step of $\Delta \tau=0.01 \text{ Ha}^{-1}$, which leads
to a discretization of the path integral into 100,000 slices at the lowest
temperature (when $\beta=10^{3}$ Ha$^{-1}$ for $T\approx 300$ K).

\subsection{Born-Oppenheimer energy surface}
\begin{figure}
\includegraphics[width=\linewidth]{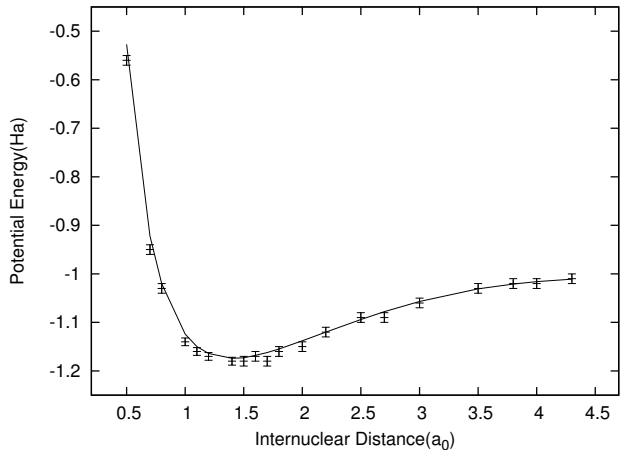} 
\caption{\label{fig:curve} Born-Oppenheimer surface of H$_2$ molecule. 
Solid line is the accurate potential energy curve from the reference 
\cite{Kolos65:1965} and data points are PIMC results using the virial energy 
estimator.}
\end{figure}

To check the accuracy of our discretized path integral,
we calculate the potential energy surface of the ground state H$_2$ molecule
with the internuclear separation from 0.5 $a_0$ to 4.5 $a_0$. 
For this calculation we fixed the proton's position of the H$_2$ molecule, 
then calculate the thermal energy using the virial energy estimator.
Figure~\ref{fig:curve} shows a very good agreement with the accurate potential energy 
curve(solid line)~\cite{Kolos65:1965}. Note that we do not use this energy surface for
any of the subsequent calculations. We only present the calculation here to verify
that we can get the correct detailed energy surface if we so desire. All following results
are directly sampled  from the electron-ion density matrix.

\subsection{Vibrational frequency of a H$_2$ molecule}
\begin{figure}
\includegraphics[width=\linewidth]{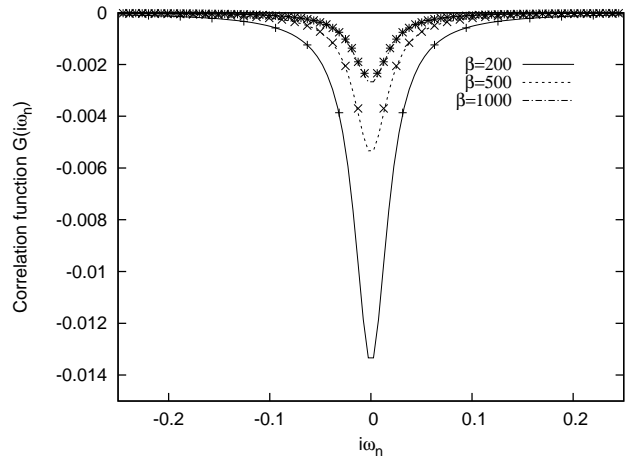} 
\caption{\label{fig:blomega} The displacement-displacement correlation functions of imaginary- 
frequency  for a hydrogen molecule at three different temperatures. 
The lines are the correlation functions of imaginary-frequency of harmonic 
oscillator at each temperatures and points represent our results before fitting 
the vibrational frequencyinto the correlation function.}
\end{figure}

By sampling the displacement-displacement correlation function in imaginary time, 
we have calculated the vibration frequencies of a hydrogen molecule (H$_2$) at three 
different inverse temperatures, $\beta=200, 500$, and $1000$ Ha$^{-1}$, 
corresponding to $T \approx 1500$K, $600$K, and $300$K.
Figure~\ref{fig:blomega} shows the correlation function in imaginary-frequency domain 
for a hydrogen molecule.
The solid lines are the displacement-displacement correlation functions of imaginary-frequency for the simple harmonic oscillator at each temperatures and the points represent our data points for the Hydrogen molecule. As described in Sec.~\ref{sec:vibration}, we extract the frequency
using two different approaches: (i) a fit to a simple harmonic oscillator 
and (ii) the full-width at half-maximum of the frequency response. 
The results are summarized in Table~\ref{table:frequency}.

\begin{table}
\caption{Vibrational frequencies calculated from PIMC simulations,
using a fit to a simple harmonic oscillator (SHO fit) and a measurement
of the full-width at half-maximum in the imaginary frequency response.}
\label{table:frequency}
\begin{ruledtabular}
\begin{tabular}{cccc}
&&\multicolumn{2}{c}{$\hbar\omega = E_1 - E_0$}\\
$\beta(\text{Ha}^{-1})$ & T(K) & SHO fit & FWHM\\
\hline
200 & 1500 & 0.01866(5) & 0.01867\\
500 &  600 & 0.01808(5) & 0.01810 \\
1000 & 300 & 0.01786((5) & 0.01788\\
\end{tabular}
\end{ruledtabular}
\end{table}

The exact ground state vibrational frequency of H$_2$ is 0.02005 in atomic 
units (=4401.21 cm$^{-1}$) \cite{Huber:1979}. Due to the anharmonic properties of
the hydrogen molecule's potential surface, the difference between the ground state and
the first excite state level is 0.01895 hartree (4161.14 cm$^{-1}$) \cite{Wolniewicz:1993}.
The time correlation functions measure this physical energy difference, not the unphysical exact ground state frequency.

It is shown that the frequency is not sensitive to the temperatures.
Because the thermal energy is so small, it just affects the rotational levels.
Since we doesn't fix the protons in the space, 
there are rotational effects to the calculation of frequency.
The analytical results (solid lines in Fig.~\ref{fig:blomega}) 
are based on the harmonic oscillator without rotational motion.

\subsection{Bond-length of a H$_2$ molecule}
\begin{figure}
\includegraphics[width=\linewidth]{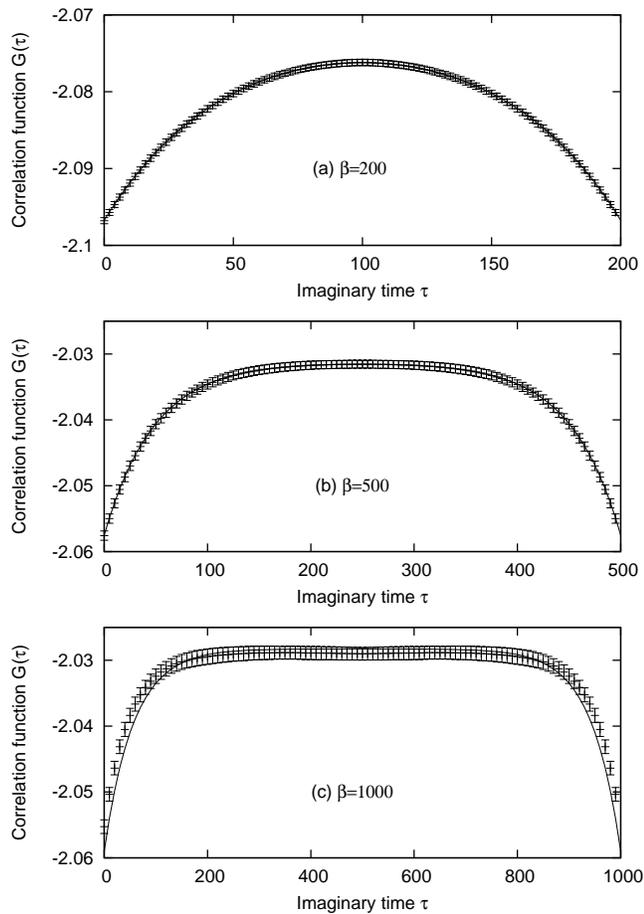} 
\caption{\label{fig:h2bl} The displacement-displacement correlation function of imaginary time 
for a hydrogen molecule at three different temperatures. 
The solid lines are analytical correlation function of simple harmonic 
oscillator and the points are PIMC results at each temperatures.}
\end{figure}

Next, we calculate the bond length of the H$_2$ molecule from imaginary-time 
correlation function. Figure~\ref{fig:h2bl} shows the displacement-displacement 
correlation functions of imaginary time at three temperatures.
At room temprature, Fig.~\ref{fig:h2bl} (c), we could oberve the flat in the middle of the imaginary time,
which shows the corrleation function is fully uncorrelated.
  
The exact bond length of a ground state hydrogen molecule is 1.401 in atomic units \cite{Huber:1979}. 
With bond-length estimator we directly calculate two average values of the 
internuclear distance, $\langle D \rangle$ and $\langle D^2 \rangle $. 
Table~\ref{tab:bl} shows the summary of two bond length averages 
from the estimator and the correlation function of a hydrogen molecule
 in atomic units. It shows the two different methods agree very 
well. 

The thermal effects on the bond-length of a molecule can be calculated 
from the Boltzmann factor $(2J+1)e^{-\beta E_J}$  with $2J+1$-fold degeneracy 
 due to the rotational motion, where $J$ is the total angular momentum.
 The energy is $E_J=BJ(J+1)$, where $B$ is the rotational constants of a hydrogen molecule, $B=60.853 \text{ cm}^{-1}$. 
At temperature, $k_BT=1/\beta$, the number of molecules $N_J$ in the rotational level $J$
can be calculated by 
\begin{equation}
N_J=\frac{(2J+1)e^{-\beta BJ(J+1) } } {\sum_J (2J+1)e^{-\beta BJ(J+1)}}.
\end{equation}
With the number of distribution with respect to the temperature, 
we can calculate the effective potential  due to the rotational motion in explicit method using 
Morse potential.
We observed that the bond-length has changed the factor of $10^{-2}$.

\begin{table}
\caption{\label{tab:bl} The summary of two bond length averages from the bond-length
estimator and the displacement-displacement correlation function of a hydrogen molecule (H$_2$) by path integral simulation (PIMC CF) in atomic units. } 
\begin{center}
 \begin{ruledtabular}
 \begin{tabular}{cccccc}
       &  &\multicolumn{2}{c}{Bond-length estimator }   & \multicolumn{2}{c}  {PIMC CF}        \\
 $\beta (\text{Ha}^{-1})$   & T(K) &$\langle D \rangle ^2$ &  $\langle D^2 \rangle $ & at  $\tau=\beta/2 $ & at $\tau=0$ \\ \hline
        200  &  1500 &2.070(4)   &   2.101(5)  &  2.076(1)  & 2.097(1)   \\
        500  &  600 &2.029(4)   &   2.056(5)  &  2.031(1)   & 2.057(1)    \\
       1000 &  300 &2.024(4)   &   2.051(5)  &  2.029(1)   & 2.055(1)   \\ 
  \end{tabular}
 \end{ruledtabular}
 \end{center}
\end{table}

\subsection{Polarization}
Many-body perturbation methods are usually used for calculating the    
polarizability of molecules. In the case of the hydrogen molecule, the 
wavefunctions are very well known so that the variational perturbation methods
of the polarizability of the molecule agree very well with the experimental
results \cite{Ishiguro:1951, Kolos67:1967}.
With the same method the polarizability have been calculated of the depedence
 on rotation and vibration states \cite{Rychlewski:1980} 
and of the function of internuclear separation \cite{Hyams:1994}.

We have calculated the polarizability of a hydrogen molecule in PIMC using two different approaches.
The first method is to use the polarization estimator as we explained in the 
estimator section, see Eq~(\ref{eq:statpol}). 
By applying a small electric field through the z-axis to a fixed molecule,
we have calculated the polarzability at room temperature, 
for the parallel direction $\alpha_\parallel$ = 6.12(50), and 
for perpendicular $\alpha_\perp$= 4.92(26) in atomic units.
The exact value is $\alpha_\parallel$=6.380 and $\alpha_\perp$=4.577 \cite{Kolos67:1967}.

\begin{figure}
  \includegraphics[width=\linewidth]{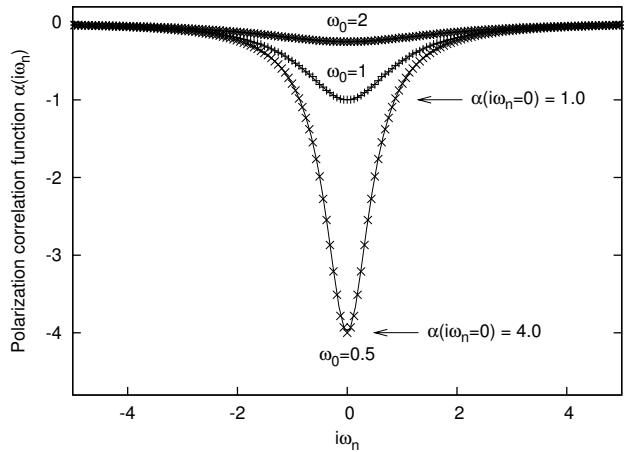} 
  \caption{\label{fig:shopol} The polarization-polarization correlation function 
    for simple harmonic oscillators with three different harmonic strength, 
    $\omega_{\mathtt{SHO}}=0.5, 1, 2$. At $i\omega_n=0$, the arrows show 
    the polarizability for each oscillators.}
\end{figure}

Another way to calculate the polarizability is to use the polarization correlation function.  
We calculate the polarizability of the simple harmonic oscillator
for three different oscillator strength $\omega_{\mathtt{SHO}}=0.5, 1, 2$ 
for the test of the correlation function method.
As we mention in the estimator section, the static polarizability can be 
obtained at $i\omega_n=0$ of $\alpha(i\omega_n)$.
As the harmonic strength increase, the polarizability becomes 
smaller, see the equation~(\ref{eq:SHOpol}). 
Figure~\ref{fig:shopol} shows that the values of polarizability for the different 
harmonic strengths exactly follow Eq.~(\ref{eq:SHOpol}). 
The polarizabilities for each oscillators are for $ \omega_{\mathtt{SHO}}=1$, 
$\alpha=1$, for $ \omega_{\mathtt{SHO}}=0.5$, $\alpha=4$ 
and for $ \omega_{\mathtt{SHO}}=2$, $\alpha=0.25$ respectively at $i\omega_n=0$.

We can also estimate the possible dipole transition energy levels ($\hbar\omega_{\mathtt{SHO}}$)
from the values of the full width at half maximum (FWHM) of correlation function.
With the analytic continuation we can obtain retarded Green's function ($G^R(\omega)$) 
of real frequency ($\omega$) from the imaginary-frequency correlation function ($G(i\omega_n)$).
The retarded Green's function is also related the spectral density function ($A(\omega$))
for the harmonic oscillator at $\omega>0$,
\begin{equation}
  \begin{aligned}
G^R(\omega) &= \frac{1}{2\pi}\int_{-\infty}^{\infty} d\omega' \frac{A(\omega')}{\omega-\omega'-i\eta} \\
            &= \frac{1/m}{(\omega+i\eta)^2-\omega_\mathtt{SHO}^2}    .
  \end{aligned}
\end{equation}

At the poles $\omega=\omega_\mathtt{SHO}$, the spectral function is a delta function
and imaginary-frequency correlation function form a Lorentian with FWHM of $\omega_\mathtt{SHO}$.
Since the harmonic oscillator has one possible transition,
the values of FWHM are exactly the same as the harmonic strengths.
But for a atom and molecule the FWHM doesn't represent the possible transition levels,
because the levels are not equivalent.

\begin{figure}
\includegraphics[width=\linewidth]{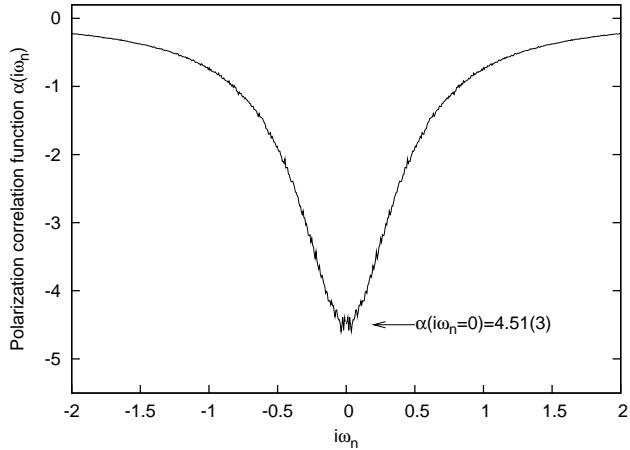} 
\caption{\label{fig:hatompol} The polarization-polarization correlation function 
of a hydrogen atom. At $i\omega_n=0$, the arrow shows the polarzability of a hydrogen atom, 4.51(3). 
The exact value is 4.5 in atomic units.}
\end{figure}

Next we test the correlation function method for calculating the polarizability of a hydrogen 
atom. We found that the polarization-polarzation correlation function has the same form 
whether the proton of the atom is fixed or not. 
Figure~\ref{fig:hatompol} shows the correlation function at $\beta=1000 \text{ Ha}^{-1}$.
 We have obtained the polarizability is  4.51(3) in atomic units, while 
 the exact value is 4.5. 

For the hydrogen molecule, we fixed the positions of two protons at the z-axis,
so that the rotation effect doesn't need to be considered for the polarization of the molecule
and thus the correlation function is independent of the temperature.

\begin{figure}
\includegraphics[width=\linewidth]{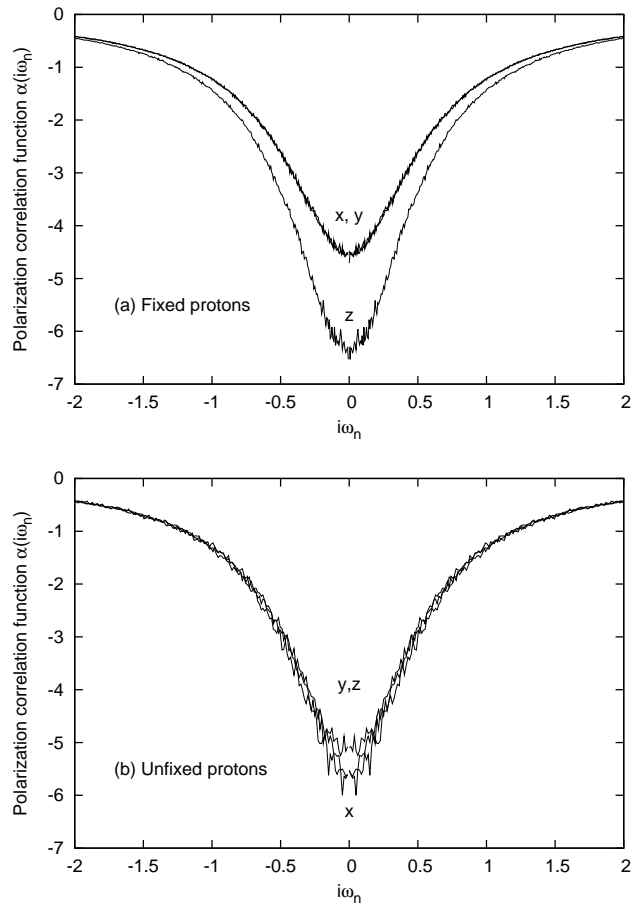} 
\caption{\label{fig:h2molpolb1000} The polarizability of a hydrogen molecule at 
$\beta=1000$. The protons are fixed on the z-axis for the upper panel 
and unfixed at the lower panel. The x, y components and 
the z component of polarizability represent $\alpha_\perp$ and $\alpha_\parallel$ respectively.}
\end{figure}

Figure~\ref{fig:h2molpolb1000} shows the imginary-frequency correlation function of a hydrogen molecule
when the protons are fixed (a) and unfixed (b).
For the perpendicular ($\alpha_\perp$) and parallel ($\alpha_\parallel$) polarizability we need to
calculate x,y and z components of the displacement of particles(two electrons and two protons) respectively.
We have obtained the perpendicular $\alpha_\perp=4.56(17)$ and 
parallel $\alpha_\parallel=6.40(15)$ polarizability.   

The value of FWHM is 0.596 for the perpendicular ($\alpha_\perp$) and 0.534 for the 
parallel ($\alpha_\parallel$) of the correlation function.
The polarizability of a molecule is a physical quantity how the molecule induced by
an electric field.
For a diatomic mole possible dipole transition energy levels for ($\alpha_\perp$) is bigger than
($\alpha_\parallel$) because of the molecule structure.

We also have calculated the correlation function with unfixed protons, allowing rotational and vibrational motion of the molecule. 
Since the trace of the polarizability tensor is invariant, the average polarizability,
$\alpha=\dfrac{1}{3}(2\alpha_\perp+\alpha_\parallel)$, 
should be identical with different configuration, see Fig.~\ref{fig:h2molpolb1000} (b).
The exact average polarzability is 5.4139 in atomic units \cite{Kolos67:1967}. 
This is the same value of nonfixed protons.
Table~\ref{tab:pol} shows the summary of the polarizability for the simple 
harmonic oscillators, the hydrogen atom and the hydrogen molecule.

\begin{table}
\begin{center}
\caption{\label{tab:pol} The summary of the polarizability of simple harmonic 
oscillators (SHO), a hydrogen atom (H), and a hydrogen molecule (H$_2$) 
for the fixed and unfixed protons using path integral Monte Carlo method with polarization-polarization 
correlation function methods (PIMC-CF) in atomic units.
The exact data of H$_2$ are from reference \cite{Kolos67:1967}.}
 \begin{ruledtabular}\begin{tabular}{cccc}
      & $\omega_{\mathtt{SHO}}$   & PIMC-CF    & Exact    \\ \hline
SHO    &     0.5    &  4.00(6)   & 4   \\
       &      1     &   1.00(6)  & 1         \\
       &      2     &   0.25(6)  & 0.25     \\ 
       &&& \\
H atom &    $\alpha$      &   4.51(3)  & 4.5       \\
H$_2$ (fixed) & $ \alpha_\perp$        &  4.56(17)   &   4.577   \\
              & $\alpha_\parallel$    &  6.40(15)    &   6.380   \\
H$_2$(unfixed)& $\alpha$      & 5.38(18)    &    5.413   \\
\end{tabular}
\end{ruledtabular}
\end{center}
\end{table}

\section{Conclusions}
We have demonstrated analysis techniques for path integral Monte Carlo
simulations of molecules. The aim of these calculations is to show how
important physical properties of molecules---energies, bond lengths,
vibrational frequencies, and polarizability---can be computed directly from
a path integral simulation without an explicit calculation of energies on 
the Born-Openheimer surface. This paper has focused on the analysis of
Matsubura  Green's functions for bond length and electrical dipole operators.
We have limited this analysis to the H$_2$ molecule to avoid complications
from fixed-node approximations. Future calculations will need to properly
treat fermions, most likely with a fixed-node approximation, if these techniques
are to have use in practical molecular simulations. The results presented
here are benchmarks that validate continued development of the PIMC technique.

We have two main findings: (1) displacement-displacement Green's functions
can give accurate estimates of the energy splitting $\hbar\omega$ between
the ground and first-excited vibrational states, and (2) dipole-dipole Green's
functions provide accurate estimates of the electrical polarizability of atoms and
molecules. Implicit in these results is the demonstration of very high accuracy
of the discretized path integral through the use of a carefully tabulated coulomb
action.

It is important to the computational strategy we are advocating in these calculations:
rather than using costly many-body techniques to calculate individual points
on the Born-Oppenheimer surface, we are including the quantum and thermal average 
of ionic positions in the many-body electronic calculation. This approach may
lead to better algorithmic efficiency for systems that have significant thermal
or zero-point ionic motion, including systems with strong polaronic effects. The merits of 
this approach relative to other QMC techniques, such as CEIMC, cannot be
measured until we treat larger molecular systems. More complicated molecules
will also lead to more complicated Green's functions that encode information on
the normal modes. Nevertheless, the benchmark calculations presented here
are a necessary first step towards future calculations on more interesting molecules.

\begin{acknowledgments}
Work supported by NSF Grant No.\ DMR 0239819 and 
made use of facilities provided by the Ira A. Fulton High Performance Computing Initiative.
\end{acknowledgments}


\begin{thebibliography}{33}

\expandafter\ifx\csname natexlab\endcsname\relax\def\natexlab#1{#1}\fi
\expandafter\ifx\csname bibnamefont\endcsname\relax
  \def\bibnamefont#1{#1}\fi
\expandafter\ifx\csname bibfnamefont\endcsname\relax
  \def\bibfnamefont#1{#1}\fi
\expandafter\ifx\csname citenamefont\endcsname\relax
  \def\citenamefont#1{#1}\fi
\expandafter\ifx\csname url\endcsname\relax
  \def\url#1{\texttt{#1}}\fi
\expandafter\ifx\csname urlprefix\endcsname\relax\def\urlprefix{URL }\fi
\providecommand{\bibinfo}[2]{#2}
\providecommand{\eprint}[2][]{\url{#2}}

\bibitem[{\citenamefont{Guardiola}(1998)}]{Guardiola:1998}
\bibinfo{author}{\bibfnamefont{R.}~\bibnamefont{Guardiola}}, in
  \emph{\bibinfo{booktitle}{Microscopic Quantum Many-Body Theories and Their
  Applications}}, edited by
  \bibinfo{editor}{\bibfnamefont{J.}~\bibnamefont{Navarro}} \bibnamefont{and}
  \bibinfo{editor}{\bibfnamefont{A.}~\bibnamefont{Polls}}
  (\bibinfo{publisher}{Springer}, \bibinfo{year}{1998}), pp.
  \bibinfo{pages}{269--336}.

\bibitem[{\citenamefont{Reynolds et~al.}(1982)\citenamefont{Reynolds, Ceperley,
  Alder, and William A.~Lester}}]{Reynolds:1982}
\bibinfo{author}{\bibfnamefont{P.~J.} \bibnamefont{Reynolds}},
  \bibinfo{author}{\bibfnamefont{D.~M.} \bibnamefont{Ceperley}},
  \bibinfo{author}{\bibfnamefont{B.~J.} \bibnamefont{Alder}}, \bibnamefont{and}
  \bibinfo{author}{\bibfnamefont{J.}~\bibnamefont{William A.~Lester}},
  \bibinfo{journal}{J. Chem. Phys.} \textbf{\bibinfo{volume}{77}},
  \bibinfo{pages}{5593} (\bibinfo{year}{1982}).

\bibitem[{\citenamefont{Hammond et~al.}(1994)\citenamefont{Hammond,
  W.~A.~Lester, and Reynolds}}]{Hammond:1994}
\bibinfo{author}{\bibfnamefont{B.~L.} \bibnamefont{Hammond}},
  \bibinfo{author}{\bibfnamefont{J.}~\bibnamefont{W.~A.~Lester}},
  \bibnamefont{and} \bibinfo{author}{\bibfnamefont{P.~J.}
  \bibnamefont{Reynolds}}, \emph{\bibinfo{title}{Monte Carlo methods in Ab
  Initio quantum chemistry}} (\bibinfo{publisher}{World Scientific},
  \bibinfo{year}{1994}).

\bibitem[{\citenamefont{Foulkes et~al.}(2001)\citenamefont{Foulkes, Mitas,
  Needs, and Rajagopal}}]{Foulkes:2001}
\bibinfo{author}{\bibfnamefont{W.~M.~C.} \bibnamefont{Foulkes}},
  \bibinfo{author}{\bibfnamefont{L.}~\bibnamefont{Mitas}},
  \bibinfo{author}{\bibfnamefont{R.~J.} \bibnamefont{Needs}}, \bibnamefont{and}
  \bibinfo{author}{\bibfnamefont{G.}~\bibnamefont{Rajagopal}},
  \bibinfo{journal}{Rev. Mod. Phys} \textbf{\bibinfo{volume}{73}},
  \bibinfo{pages}{33} (\bibinfo{year}{2001}).

\bibitem[{\citenamefont{Shumway and Ceperley}(2006)}]{Shumway:2006b}
\bibinfo{author}{\bibfnamefont{J.}~\bibnamefont{Shumway}} \bibnamefont{and}
  \bibinfo{author}{\bibfnamefont{D.~M.} \bibnamefont{Ceperley}}, in
  \emph{\bibinfo{booktitle}{Handbook of Theoretical and Computational
  Nanotechnology}}, edited by
  \bibinfo{editor}{\bibfnamefont{M.}~\bibnamefont{Rieth}} \bibnamefont{and}
  \bibinfo{editor}{\bibfnamefont{W.}~\bibnamefont{Schommers}}
  (\bibinfo{publisher}{American Scientific Publishers},
  \bibinfo{address}{Germany}, \bibinfo{year}{2006}).

\bibitem[{\citenamefont{Grossman}(2002)}]{Grossman:2002}
\bibinfo{author}{\bibfnamefont{J.~C.} \bibnamefont{Grossman}},
  \bibinfo{journal}{J. Chem. Phys.} \textbf{\bibinfo{volume}{117}},
  \bibinfo{pages}{1434} (\bibinfo{year}{2002}).

\bibitem[{\citenamefont{Benedek et~al.}(2006)\citenamefont{Benedek, Snook,
  Towler, and Needs}}]{Benedek:2006}
\bibinfo{author}{\bibfnamefont{N.~A.} \bibnamefont{Benedek}},
  \bibinfo{author}{\bibfnamefont{I.~K.} \bibnamefont{Snook}},
  \bibinfo{author}{\bibfnamefont{M.~D.} \bibnamefont{Towler}},
  \bibnamefont{and} \bibinfo{author}{\bibfnamefont{R.~J.} \bibnamefont{Needs}},
  \bibinfo{journal}{J. Chem. Phys.} \textbf{\bibinfo{volume}{125}},
  \bibinfo{pages}{104302} (\bibinfo{year}{2006}).

\bibitem[{\citenamefont{Ceperley et~al.}(2002)\citenamefont{Ceperley, Dewing,
  and Pierleoni}}]{Ceperley:2002}
\bibinfo{author}{\bibfnamefont{D.~M.} \bibnamefont{Ceperley}},
  \bibinfo{author}{\bibfnamefont{M.}~\bibnamefont{Dewing}}, \bibnamefont{and}
  \bibinfo{author}{\bibfnamefont{C.}~\bibnamefont{Pierleoni}}, in
  \emph{\bibinfo{booktitle}{Topics in Condensed Matter Physics}}, edited by
  \bibinfo{editor}{\bibfnamefont{P.}~\bibnamefont{Nielaba}},
  \bibinfo{editor}{\bibfnamefont{M.}~\bibnamefont{Mareschal}},
  \bibnamefont{and} \bibinfo{editor}{\bibfnamefont{G.}~\bibnamefont{Ciccotti}}
  (\bibinfo{publisher}{Springer-Verlag}, \bibinfo{year}{2002}).

\bibitem[{\citenamefont{Pierleoni and Ceperley}(2006)}]{Pierleoni:2006}
\bibinfo{author}{\bibfnamefont{C.}~\bibnamefont{Pierleoni}} \bibnamefont{and}
  \bibinfo{author}{\bibfnamefont{D.~M.} \bibnamefont{Ceperley}},
  \bibinfo{journal}{ChemPhysChem} \textbf{\bibinfo{volume}{6}},
  \bibinfo{pages}{1872} (\bibinfo{year}{2006}).

\bibitem[{\citenamefont{Shumway}(2005)}]{Shumway:2005d}
\bibinfo{author}{\bibfnamefont{J.}~\bibnamefont{Shumway}}, in
  \emph{\bibinfo{booktitle}{Computer Simulations Studies in Condensed Matter
  Physics}}, edited by \bibinfo{editor}{\bibfnamefont{D.~P.}
  \bibnamefont{Landau}}, \bibinfo{editor}{\bibfnamefont{S.~P.}
  \bibnamefont{Lewis}}, \bibnamefont{and} \bibinfo{editor}{\bibfnamefont{H.-B.}
  \bibnamefont{{Sch\"uttler}}} (\bibinfo{publisher}{Springer},
  \bibinfo{address}{Berlin}, \bibinfo{year}{2005}), vol.
  \bibinfo{volume}{XVII}.

\bibitem[{\citenamefont{Baym and Mermin}(1961)}]{Baym:1961}
\bibinfo{author}{\bibfnamefont{G.}~\bibnamefont{Baym}} \bibnamefont{and}
  \bibinfo{author}{\bibfnamefont{N.~D.} \bibnamefont{Mermin}},
  \bibinfo{journal}{J. Math. Phys.} \textbf{\bibinfo{volume}{2}},
  \bibinfo{pages}{232} (\bibinfo{year}{1961}).

\bibitem[{\citenamefont{Thirumalai and Berne}(1983)}]{Thirumalai:1983}
\bibinfo{author}{\bibfnamefont{D.}~\bibnamefont{Thirumalai}} \bibnamefont{and}
  \bibinfo{author}{\bibfnamefont{B.~J.} \bibnamefont{Berne}},
  \bibinfo{journal}{J. Chem. Phys.} \textbf{\bibinfo{volume}{79}},
  \bibinfo{pages}{5029} (\bibinfo{year}{1983}).

\bibitem[{\citenamefont{Schuttler and Scalapino}(1985)}]{Schuttler:1985}
\bibinfo{author}{\bibfnamefont{H.~B.} \bibnamefont{Schuttler}}
  \bibnamefont{and} \bibinfo{author}{\bibfnamefont{D.~J.}
  \bibnamefont{Scalapino}}, \bibinfo{journal}{Phys. Rev. Lett.}
  \textbf{\bibinfo{volume}{55}}, \bibinfo{pages}{1204} (\bibinfo{year}{1985}).

\bibitem[{\citenamefont{Jarrell and Biham}(1989)}]{Jarrell:1989}
\bibinfo{author}{\bibfnamefont{M.}~\bibnamefont{Jarrell}} \bibnamefont{and}
  \bibinfo{author}{\bibfnamefont{O.}~\bibnamefont{Biham}},
  \bibinfo{journal}{Phys. Rev. Lett.} \textbf{\bibinfo{volume}{63}},
  \bibinfo{pages}{2504} (\bibinfo{year}{1989}).

\bibitem[{\citenamefont{Gallicchio and Berne}(1996)}]{Gallicchio:1996}
\bibinfo{author}{\bibfnamefont{E.}~\bibnamefont{Gallicchio}} \bibnamefont{and}
  \bibinfo{author}{\bibfnamefont{B.~J.} \bibnamefont{Berne}},
  \bibinfo{journal}{J. Chem. Phys.} \textbf{\bibinfo{volume}{105}},
  \bibinfo{pages}{7064} (\bibinfo{year}{1996}).

\bibitem[{\citenamefont{Krilov et~al.}(2001)\citenamefont{Krilov, Sim, and
  J.Berne}}]{Krilov:2001}
\bibinfo{author}{\bibfnamefont{G.}~\bibnamefont{Krilov}},
  \bibinfo{author}{\bibfnamefont{E.}~\bibnamefont{Sim}}, \bibnamefont{and}
  \bibinfo{author}{\bibfnamefont{B.}~\bibnamefont{J.Berne}},
  \bibinfo{journal}{J. Chem. Phys.} \textbf{\bibinfo{volume}{114}},
  \bibinfo{pages}{1075} (\bibinfo{year}{2001}).

\bibitem[{\citenamefont{Gallicchio et~al.}(1998)\citenamefont{Gallicchio,
  Egorov, and Berne}}]{Gallicchio:1998}
\bibinfo{author}{\bibfnamefont{E.}~\bibnamefont{Gallicchio}},
  \bibinfo{author}{\bibfnamefont{S.~A.} \bibnamefont{Egorov}},
  \bibnamefont{and} \bibinfo{author}{\bibfnamefont{B.~J.} \bibnamefont{Berne}},
  \bibinfo{journal}{J. Chem. Phys.} \textbf{\bibinfo{volume}{109}},
  \bibinfo{pages}{7745} (\bibinfo{year}{1998}).

\bibitem[{\citenamefont{Blinov et~al.}(2004)\citenamefont{Blinov, Song, and
  Roy}}]{Blinov:2004}
\bibinfo{author}{\bibfnamefont{N.}~\bibnamefont{Blinov}},
  \bibinfo{author}{\bibfnamefont{X.}~\bibnamefont{Song}}, \bibnamefont{and}
  \bibinfo{author}{\bibfnamefont{P.-N.} \bibnamefont{Roy}},
  \bibinfo{journal}{J. Chem. Phys.} \textbf{\bibinfo{volume}{120}},
  \bibinfo{pages}{5916} (\bibinfo{year}{2004}).

\bibitem[{\citenamefont{Feynman}(1972)}]{Feynman:1972}
\bibinfo{author}{\bibfnamefont{P.~R.} \bibnamefont{Feynman}},
  \emph{\bibinfo{title}{Statistical Mechanics}} (\bibinfo{publisher}{Westview
  Press}, \bibinfo{year}{1972}).

\bibitem[{\citenamefont{Ceperley}(1995)}]{Ceperley:1995}
\bibinfo{author}{\bibfnamefont{D.~M.} \bibnamefont{Ceperley}},
  \bibinfo{journal}{Rev. Mod. Phys} \textbf{\bibinfo{volume}{67}},
  \bibinfo{pages}{279} (\bibinfo{year}{1995}).

\bibitem[{\citenamefont{Shin et~al.}(2006)\citenamefont{Shin, Shumway, and
  Schmidt}}]{Shumway:2006}
\bibinfo{author}{\bibfnamefont{D.}~\bibnamefont{Shin}},
  \bibinfo{author}{\bibfnamefont{J.}~\bibnamefont{Shumway}}, \bibnamefont{and}
  \bibinfo{author}{\bibfnamefont{K.}~\bibnamefont{Schmidt}}
  (\bibinfo{year}{2006}), \bibinfo{note}{in preparation to submit}.

\bibitem[{\citenamefont{Hostler and Pratt}(1963)}]{Hostler:1963}
\bibinfo{author}{\bibfnamefont{L.}~\bibnamefont{Hostler}} \bibnamefont{and}
  \bibinfo{author}{\bibfnamefont{R.}~\bibnamefont{Pratt}},
  \bibinfo{journal}{Phys. Rev. Lett.} \textbf{\bibinfo{volume}{10}},
  \bibinfo{pages}{469} (\bibinfo{year}{1963}).

\bibitem[{\citenamefont{Hostler}(1970)}]{Hostler:1970}
\bibinfo{author}{\bibfnamefont{L.}~\bibnamefont{Hostler}}, \bibinfo{journal}{J.
  Math. Phys.} \textbf{\bibinfo{volume}{11}}, \bibinfo{pages}{2966}
  (\bibinfo{year}{1970}).

\bibitem[{\citenamefont{Herman et~al.}(1982)\citenamefont{Herman, Bruskin, and
  Berne}}]{Herman:1982}
\bibinfo{author}{\bibfnamefont{M.~F.} \bibnamefont{Herman}},
  \bibinfo{author}{\bibfnamefont{E.~J.} \bibnamefont{Bruskin}},
  \bibnamefont{and} \bibinfo{author}{\bibfnamefont{B.~J.} \bibnamefont{Berne}},
  \bibinfo{journal}{J. Chem. Phys.} \textbf{\bibinfo{volume}{76}},
  \bibinfo{pages}{5150} (\bibinfo{year}{1982}).

\bibitem[{\citenamefont{Mahan}(2000)}]{Mahan:2000}
\bibinfo{author}{\bibfnamefont{G.~D.} \bibnamefont{Mahan}},
  \emph{\bibinfo{title}{Many-Particle Physics}} (\bibinfo{publisher}{Kluwer
  Academic/Plenum Publishers}, \bibinfo{year}{2000}).

\bibitem[{\citenamefont{Doniach and Sondheimer}(1974)}]{Doniach:1974}
\bibinfo{author}{\bibfnamefont{S.}~\bibnamefont{Doniach}} \bibnamefont{and}
  \bibinfo{author}{\bibfnamefont{E.~H.} \bibnamefont{Sondheimer}},
  \emph{\bibinfo{title}{Green's Functions for Solid State Physicists}}
  (\bibinfo{publisher}{W. A. Benjamin, Inc.}, \bibinfo{year}{1974}).

\bibitem[{\citenamefont{Kolos and Wolniewicz}(1965)}]{Kolos65:1965}
\bibinfo{author}{\bibfnamefont{W.}~\bibnamefont{Kolos}} \bibnamefont{and}
  \bibinfo{author}{\bibfnamefont{L.}~\bibnamefont{Wolniewicz}},
  \bibinfo{journal}{J. Chem. Phys.} \textbf{\bibinfo{volume}{43}},
  \bibinfo{pages}{2429} (\bibinfo{year}{1965}).

\bibitem[{\citenamefont{Huber and Herzberg}(1979)}]{Huber:1979}
\bibinfo{author}{\bibfnamefont{K.~P.} \bibnamefont{Huber}} \bibnamefont{and}
  \bibinfo{author}{\bibfnamefont{G.}~\bibnamefont{Herzberg}},
  \emph{\bibinfo{title}{Molecular Spectra and Molecular Structure}}, vol.
  \bibinfo{volume}{4 Constants of Diatomic Molecules} (\bibinfo{publisher}{Van
  Nostrand Reinhold, New York}, \bibinfo{year}{1979}).

\bibitem[{\citenamefont{Wolniewicz}(1993)}]{Wolniewicz:1993}
\bibinfo{author}{\bibfnamefont{L.}~\bibnamefont{Wolniewicz}},
  \bibinfo{journal}{J. Chem. Phys.} \textbf{\bibinfo{volume}{99}},
  \bibinfo{pages}{1851} (\bibinfo{year}{1993}).

\bibitem[{\citenamefont{Ishiguro et~al.}(1952)\citenamefont{Ishiguro, Arai,
  Mizushima, and Kotani}}]{Ishiguro:1951}
\bibinfo{author}{\bibfnamefont{E.}~\bibnamefont{Ishiguro}},
  \bibinfo{author}{\bibfnamefont{T.}~\bibnamefont{Arai}},
  \bibinfo{author}{\bibfnamefont{M.}~\bibnamefont{Mizushima}},
  \bibnamefont{and} \bibinfo{author}{\bibfnamefont{M.}~\bibnamefont{Kotani}},
  \bibinfo{journal}{Proc. Phys. Soc. A} \textbf{\bibinfo{volume}{65}},
  \bibinfo{pages}{178} (\bibinfo{year}{1952}).

\bibitem[{\citenamefont{Kolos and Wolniewicz}(1967)}]{Kolos67:1967}
\bibinfo{author}{\bibfnamefont{W.}~\bibnamefont{Kolos}} \bibnamefont{and}
  \bibinfo{author}{\bibfnamefont{L.}~\bibnamefont{Wolniewicz}},
  \bibinfo{journal}{J. Chem. Phys.} \textbf{\bibinfo{volume}{46}},
  \bibinfo{pages}{1426} (\bibinfo{year}{1967}).

\bibitem[{\citenamefont{Rychlewski}(1980)}]{Rychlewski:1980}
\bibinfo{author}{\bibfnamefont{J.}~\bibnamefont{Rychlewski}},
  \bibinfo{journal}{Mol. Phy.} \textbf{\bibinfo{volume}{41}},
  \bibinfo{pages}{833} (\bibinfo{year}{1980}).

\bibitem[{\citenamefont{Hyams et~al.}(1994)\citenamefont{Hyams, Gerratt,
  Cooper, and Raimondi}}]{Hyams:1994}
\bibinfo{author}{\bibfnamefont{P.~A.} \bibnamefont{Hyams}},
  \bibinfo{author}{\bibfnamefont{J.}~\bibnamefont{Gerratt}},
  \bibinfo{author}{\bibfnamefont{D.~L.} \bibnamefont{Cooper}},
  \bibnamefont{and} \bibinfo{author}{\bibfnamefont{M.}~\bibnamefont{Raimondi}},
  \bibinfo{journal}{J. Chem. Phys.} \textbf{\bibinfo{volume}{100}},
  \bibinfo{pages}{4417} (\bibinfo{year}{1994}).

\end{thebibliography}
\end{document}